# Equity Market Price Changes Are Predictable: A Natural Science Approach

Qingyuan Han
University of Alabama in Huntsville
hanq@uah.edu

## Abstract

Equity markets have long been regarded as unpredictable, with intraday price movements treated as stochastic noise. This study challenges that view by introducing the Extended Samuelson Model (ESM), a natural science–based framework that captures the dynamic, causal processes underlying market behavior. ESM identifies peaks, troughs, and turning points across multiple timescales and demonstrates temporal compatibility: finer timeframes contain all signals of broader ones while offering sharper directional guidance. Beyond theory, ESM translates into practical trading strategies. During intraday sessions, it reliably anticipates short-term reversals and longer-term trends, even under the influence of breaking news. Its eight market states and six directional signals provide actionable guardrails for traders, enabling consistent profit opportunities. Notably, even during calm periods, ESM can capture 10-point swings in the S&P 500—equivalent to $500 per E-mini futures contract. These findings resonate with the state-based approaches attributed to Renaissance Technologies' Medallion Fund, which delivered extraordinary returns through systematic intraday trading. By bridging normal conditions with crisis dynamics, ESM not only advances the scientific understanding of market evolution but also provides a robust, actionable roadmap for profitable trading.

## Introduction

Market crises have a wide-range and enduring impact on both the economy and society. Economically, they trigger sharp declines in asset prices, erode household wealth, and disrupt credit markets, often leading to widespread business failures and surging unemployment (Muddarajaiah 2023). Governments face mounting fiscal pressures as they deploy emergency interventions, which can result in long-term debt accumulation and constrained public spending (Jain 2025). Socially, crises exacerbate inequality and poverty, disproportionately affecting vulnerable populations and undermining social mobility (Lei et al. 2023). They contribute to psychological distress, civil unrest, and declining trust in institutions, while political landscapes may shift dramatically as populist movements gain traction in response to perceived systemic failures (Funke et al. 2016).

Despite the efficient market hypothesis's assertion that market crises are inherently unpredictable, as all information is already in the price (Fama 1970), the devastating economic and social consequences of reccuring market collapses have fueled extensive research into early warning systems (EWS). Traditional predictors include the Credit-to-GDP Gap (Borio and Lowe 2002, Schularick and Taylor 2012), the CAPE ratio (Shiller, 2000), External Imbalances (Reinhart and Reinhart 2008), the VIX (Whaley 2000), Credit Spreads (Gorton and Metrick 2012), and Yield Curve Inversion (Estrella and Mishkin 1998). More recent approaches employ Machine Learning (Sirignano et al. 2018) and Network Analysis (Brownlees and Engle 2017). However, traditional early warning systems operate primarily as statistical pattern-recognition tools, in contrast to the causal, dynamic models prevalent in the natural sciences. The EWSs excel at flagging vulnerabilities that resemble past crises but often falter when the underlying mechanisms evolve (Borio and Drehmann 2009).

A key limitation of traditional early warning systems (EWS) lies in their reliance on statistical methods that not only struggle with out-of-sample accuracy but also oversimplify crises into binary outcomes—either a crisis or no crisis at all. This binary framing overlooks the continuous dynamics that carry markets from stability toward extreme stress. As Ragnar Frisch observed (1929/1992), statistical analysis treats events as independent, discarding sequential dynamics, akin to shuffling cards after a game: the final summary remains identical, regardless of the order of play. It has long been clear that traditional economic approaches are insufficient and new models are urgently needed. As Barberis and Thaler (2003) stated, most current theories are likely wrong and will be replaced by better ones. Farmer and Foley (2009) took this further, arguing for a complete shift to dynamic models, similar to a climate model, that capture how agents interact. Yet, moving beyond traditional ways of thinking is the biggest hurdle. This challenge was captured by John Keynes (1946), who pointed out that the real struggle is not with the new ideas, but with letting go of the old ones that are deeply embedded in our minds.

The power of Farmer and Foley's argument lies in its call to ground economics in the methodological principles of the natural sciences. Both meteorological and economic systems are characterized by extreme complexity arising from numerous interacting elements. However, a fundamental divergence is evident: meteorology, as a natural science, generates falsifiable and accountable forecasts, whereas prevailing economic paradigms frequently reject the feasibility of predicting markets. This methodological gap highlights the promise of a scientific turn in economics and prompts two central inquiries: first, what are the core methodological distinctions between economics and the natural sciences, and second, under what conditions could scientific methods be viably applied to the economic domain?

To delineate the fundamental methodological distinction, one must contrast the dynamic foundations of the natural sciences with the predominantly statistical approaches of financial economics. In disciplines such as physics and meteorology, theories are expressed through dynamic equations (e.g., Newton's laws, Kepler's laws, Navier-Stokes equations) that explicitly represent the causal mechanisms governing a system. These are not merely correlational but are constitutive, specifying how inputs produce outputs. Furthermore, the critical variables in these

models—mass, velocity, pressure—are amenable to real-time observation and measurement, thereby enabling persistent empirical validation and refinement.

Financial economic theory, in contrast, is built upon a fundamentally different epistemological basis. Its models center on the inherently unmeasurable behaviors—rational or otherwise—of market participants, relying on statistical correlations to account for price fluctuations. These approaches often rest on the assumption of independent price changes and the stability of statistical parameters over time. Although they offer predictive utility within a specific scope, their correlative nature means they do not capture the causal structure of market dynamics. This fundamental limitation renders them particularly inadequate during episodes of extreme volatility, such as financial bubbles and crashes.

This methodological contrast frames the second question concerning the feasibility of applying a natural science paradigm to finance. The requisite path forward is the construction of causal dynamic models grounded in variables that are instantaneously observable within financial markets. Such models would transcend mere statistical extrapolation by elucidating the generative mechanisms of market evolution. This would enable the modeling of how micro-level agent interactions precipitate a phase transition from stability to severe stress, thereby facilitating the early identification of systemic crises.

The theoretical groundwork for this approach was established by Paul Samuelson (1941), who formulated a dynamic model linking the excess demand of liquidity demanders to the rate of price change. Subsequent contributions by Tibor Scitovsky (1952) and Kenneth Arrow (1959) argued for the integral role of liquidity providers as price makers. However, the empirical intractability of these variables at the time ultimately impeded the theory's development.

The pioneering work of earlier economists has been revitalized by the recent capacity to measure real-time excess demand, embodied in the Extended Samuelson Model (ESM) (Han and Keen 2021, Han 2025). Functioning as a causal dynamic model, the ESM provides an explanation for daily market returns and demonstrates predictive power for major systemic collapses, such as the crises of 2000, 2008, and 2020. The model's architecture categorizes market conditions into eight discrete states with specific transition thresholds. This structure enables the observation of a market's progression from a stable regime toward a critical tipping point that precedes a major downturn. A notable illustration of this is the ESM's signal on October 9, 1987—ten days before Black Monday—which indicated the market was poised for a catastrophic shift. The downturn commenced the following trading day, ultimately erasing 27.7% of the market's value in just four days.

The failure to establish a dependable early warning system for market crises is compounded by the conventional treatment of short-term fluctuations as stochastic and thus inherently unpredictable (Shiller 2014). This paper challenges that paradigm, presenting evidence that a methodology grounded in the natural sciences yields consistent predictive insights across temporal horizons, from intra-minute to monthly intervals. Moreover, these high-frequency forecasts do not merely reflect longer-horizon signals; they generate additional, unique directional guidance, thereby unlocking new avenues for real-time financial surveillance.

The remainder of the paper is structured as follows. We begin by introducing the ESM model, detailing its six directional price signals and eight market states along with their expected transition points. The following section, Predicting Price Movements and Market Turns, illustrates how ESM can be applied to forecast market behavior across various time scales, including daily trends, intraday fluctuations, and the model's temporal consistency. Next, the News Impact on Price Changes section demonstrates that news affects prices through shaping investor expectations, rather than causing immediate adjustments to an intrinsic value—a concept that remains ambiguous even among experts. Finally, the Discussion and Conclusion section explores applications of the ESM and outlines potential directions for future research.

**Extended Samuelson Model (ESM)**

The extended Samuelson model (ESM) is expressed as (for a comprehensive understanding of the ESM's history and formation, refer to Han and Keen 2021 and Han 2025),

$$\frac{dln(p)}{dt} = H\left[\frac{D(p) - S(p)}{D(p) + S(p)}\right] + M$$

where p denotes the price, *D(p)* and *S(p)* represents demand and supply, respectively. The difference, *D(p)–S(p),* indicates the excess demand. *H* is an increasing function of excess demand (*H'>0*) and remains zero when demand and supply are balanced (*H(0)=0*). $\frac{D(p)-S(p)}{D(p)+S(p)}$ is the Normalized Excess Demand (NED) describing the aggregate behavior of price-takers, and *M* is the behavior and the market microstructure impact of price makers. Han and Keen (2021) computed the NED across six timeframes: 5-minute, 15-minute, hourly, daily, weekly, and monthly. The intraday NED calculations (5-minute, 15-minute, and hourly) were based on data from January 2, 2013, up to the present. Their findings show that intraday NED, which represents the actions of liquidity takers, accounts for a significant portion (64%, *p = 0*) of daily return variances over a ten-year period. The residual variances are, according to the extended Samuelson model, due to the behaviors of liquidity providers.

Han and Keen (2021) empirically validated their Net Excess Demand (NED) calculations through a two-month, real-time forecasting experiment using 15-minute data. The theoretical premise hinges on the liquidity provision mechanism: market makers, constrained by capital, must unwind accumulated inventories at a price above their acquisition cost, thereby inducing price reversals (Garman 1976, Madhavan and Smidt 1993). The experiment tested whether NED data could reliably infer these inventory levels to forecast reversals. The validation was highly successful: 90.7% of predicted reversals (49 of 54) materialized within the 7-day forecast horizon, a result later refined by Han (2025), who showed that using 5-minute data could elevate the success rate to 95%. The accuracy of the NED metric is thus confirmed not only by this real-time test but also by its ability to generate six distinct price-directional signals and eight market states, detailed in the following section.

**Price direction signals**

The market trajectory is shaped by the convergence or divergence of expectations between liquidity providers and demanders. The six directional signals identified by Han and Keen (2021) and Han (2025) are:

1. **Uptrend Signal**: Peaks and troughs of NED rise alongside stock price highs and lows, reflecting widespread optimism.
2. **Downtrend Signal**: NED peaks and troughs decline with price, indicating prevailing pessimism..
3. **Reverse to uptrend signal**. During a downtrend, NED forms a trough equal to or lower than the previous, while price forms a higher low, suggesting optimism among liquidity providers.
4. **Reverse to downtrend signal**. During an uptrend, NED peaks match or exceed prior peaks, but price peaks fall below previous highs, signaling potential downtrend.
5. **Selling signal from informed liquidity takers.** NED drops at a market high or below previous peaks, indicating profit-taking.
6. **Buying signal from infromed liquidity takers.** NED rises at a market low or above prior lows, signaling potential rebounds.

These signals have successfully anticipated major crises in 1987, 2000, 2008, and 2020.

**Market states and Expected Transition Points**

Financial markets are dynamic, shaped by a mixture of trading strategies that cater to various time horizons and their corresponding future expectations. This blend results in differing degrees of overall market sentiment. Han (2025) categorized these varying levels of expectation into eight distinct market states, ranging from the most pessimistic (state 1) to the most euphoric (state 8), as detailed in the following table.

Table 1 Eight market states

|  | 1 | 2 | 3 | 4 | 5 | 6 | 7 | 8 |
|---|---|---|---|---|---|---|---|---|
| **Daily NED** | N | P | N | P | N | P | N | P |
| **Weekly NED** | N | N | P | P | N | N | P | P |
| **Monthly NED** | N | N | N | N | P | P | P | P |

N: Negative, P: Positive

Analysis of 25 years of data (1999-2023) reveals a positive correlation between market states and S&P 500 levels (Han 2025). This aligns with the basic economic principle that increased buying activity drives prices higher. While this correlation isn't absolute—due to the influence of liquidity providers—it's notable that market highs consistently coincide with State 8 (most euphoric), and market troughs are always linked to State 1 (most pessimistic). These associations offer a general framework for profitable trading. As shown in Han (2025), because market states are defined by NEDs, they can also be used like NEDs to predict price directions using the six directional signals.

Besides identifying market states, the Extended Samuelson Model (ESM) also forecasts turning points between positive and negative NED across various timeframes. This capability offers a robust basis for risk assessment. As Han (2025) illustrates, ten days before Black Monday, specifically on October 9, 1987, the Extended Samuelson Model (ESM) signaled a drastic market dive from State 7 to State 1. This prediction materialized the very next trading day, initiating several days of escalating market plunges that culminated in Black Monday. The continuously forecasted turning points serve as a constant guardrail, enabling real-time assessment of the risk of a market falling and the potential for a market surging, driven by anticipated shifts in market states.

**Predicting Price Movements and Market Turns**

Han and Keen (2021) utilized the six price direction signals to predict market crises in 2000, 2008, and 2020. Han (2025) expanded on this by incorporating the eight market states and their transition points to explain how they could have forewarned the impending Black Monday crash in October 1987. Below, we'll first apply these techniques to the most recent daily stock market changes in 2025. Next, we'll show how these techniques can also predict short-term fluctuations, even down to intraday movements, across both the most volatile and calmest days in the market.

**Daily Changes**

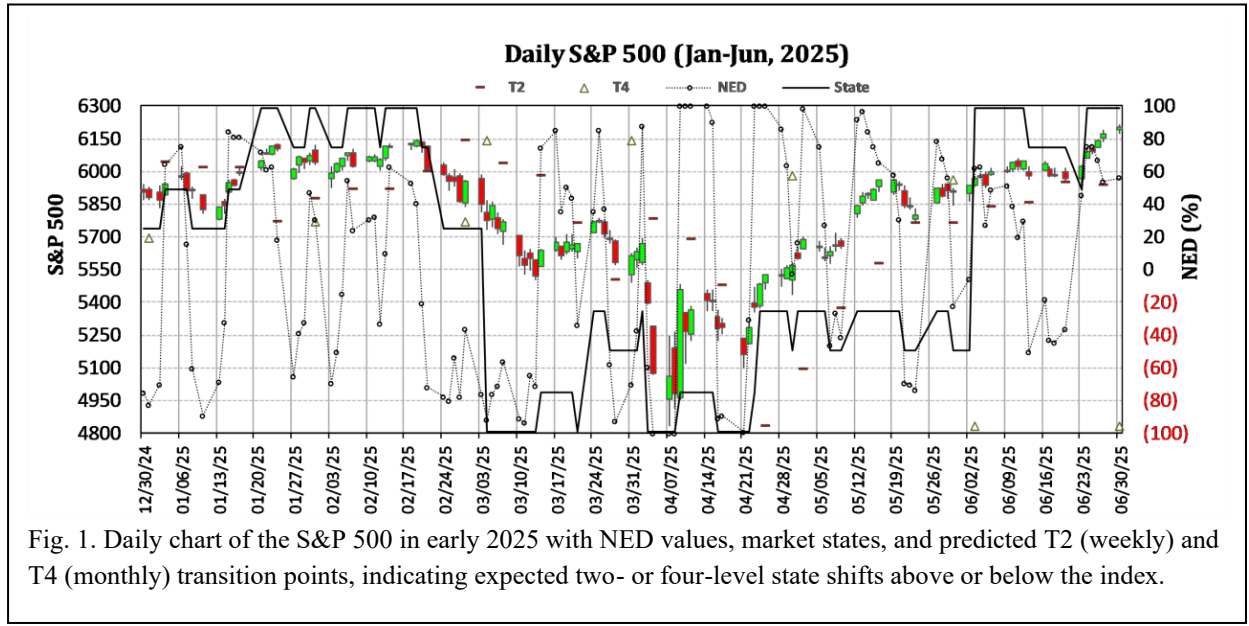

Fig. 1. Daily chart of the S&P 500 in early 2025 with NED values, market states, and predicted T2 (weekly) and T4 (monthly) transition points, indicating expected two- or four-level state shifts above or below the index.

Figure 1 illustrates that in 2025, the S&P 500 followed the established pattern: market highs corresponded to State 8, and lows to State 1. On January 20th, the market entered State 8, reflecting optimism surrounding the presidential inauguration. This early enthusiasm diminished due to slowing economic fundamentals, inflation concerns, tariff uncertainties, and company-specific negative news.

On February 18–19, despite the S&P 500 reaching new record highs, declining NED values (Signal 5) suggested informed traders were taking profits. This occurred just prior to the release of several disappointing economic data reports on February 21. Consequently, the market experienced a large sell-off on February 21, reaching 6008.56—near the upcoming weekly turning point (T2) at 6003, signaling a potential transition from State 7 to State 5. The market subsequently fell to State 5, moving toward the monthly turning point (T4) and suggesting a further decline toward State 1. This was confirmed by a Signal 4 emerging on February 28, which accurately predicted the market's drop to State 1 on March 2.
State 1 indicates a market trough but not necessarily the absolute low. Subsequent market movements depend on the position of turning points and directional signals. In early March, T4 and T2 turning points remained above the declining S&P 500, indicating limited prospects for an immediate rebound. By the third week, T2 had moved closer to the leveling market, and Signal 3 appeared on March 21, forecasting an upward shift. The market responded as predicted, transitioning from State 1 to State 4 in the fourth week.

Because the T4 (up) turning point was positioned well above the S&P 500 while the T2 (down) point lay much closer below, the market was predisposed to decline. This tendency was confirmed by a Signal 4 on April 2, indicating strong downward pressure. That evening,

President Donald Trump's "Liberation Day" declaration and announcement of sweeping tariffs further amplified selling pressure, pushing the S&P 500 down 10.5% over the next two sessions. The downturn then evolved into extreme volatility, triggered by conflicting developments: rumors of a 90-day tariff pause later denied by the White House (April 7); Trump's April 9 social media post ("THIS IS A GREAT TIME TO BUY!!! DJT") at 9:37 a.m.; his subsequent 1:18 p.m. suspension of most tariffs; the White House's clarification that Chinese tariffs had instead risen to 145%; and China's retaliatory tariffs on April 10.

Between April 9 and 15, a Signal 4 pushed the market into State 1, followed by a persistent Signal 3 through April 21 forecasting recovery. Supported by a sequence of Signals 1 and 3 and the narrowing gap between the S&P 500 and T4, this rebound carried through May. By month-end, confidence was high that the market would reach State 8 in June.

Figure 1 illustrates two critical cases where directional signals overrode conventional technical cues. On May 23, the S&P 500 reached a T2 turning point, which typically precipitates a decline from State 3 to State 1. However, the simultaneous activation of a Signal 3 propelled the market upward into State 4 instead. This demonstrates that turning points must be interpreted in conjunction with directional signals to accurately forecast state transitions. A parallel case occurred around June 20-23. Despite the market's initial expectation of a downturn following the U.S. launch of Operation Midnight Hammer on June 22, a potent Signal 3 emerged on June 23, driving the S&P 500 to new highs. These episodes underscore a fundamental principle: market direction is governed by the collective interpretation of events—reflected in these signals—rather than by the nominal positive or negative character of the news itself.

In summary, the integration of daily S&P 500 price changes with directional signals and turning points effectively identifies major peaks, troughs, and trends. However, this daily-level analysis cannot yet explain why certain days sustain strong directional moves or why intraday reversals occur. The following section will demonstrate how applying this same analytical framework to higher-frequency, intraday data can capture these very fluctuations, which have long been dismissed as random noise.

**Intraday Fluctuations**

To test short-term predictability, we compare volatile days (April 7–9) with stable ones (February 14, 18, 19). Intraday volatility $V$, defined as the ratio of the high-low range to the prior close, averaged 1.49% across 122 trading days in early 2025. On April 7–9, $V$ spiked to 8.11%, 7.05%, and 10.70%, while on February 14, 18, and 19 it was just 0.32%, 0.49%, and 0.59%. These contrasts show that intraday behavior, whether turbulent or calm, can be forecast using the same methods at finer intervals.

**Turbulent Intraday Price Movements**

Figure 2 illustrates April 7–9 using 5-minute data for S&P 500 levels, NED values, states, and T2/T4 points. States are derived from 5-, 15-, and 30-minute data, while turning points come from 15- and 30-minute data.

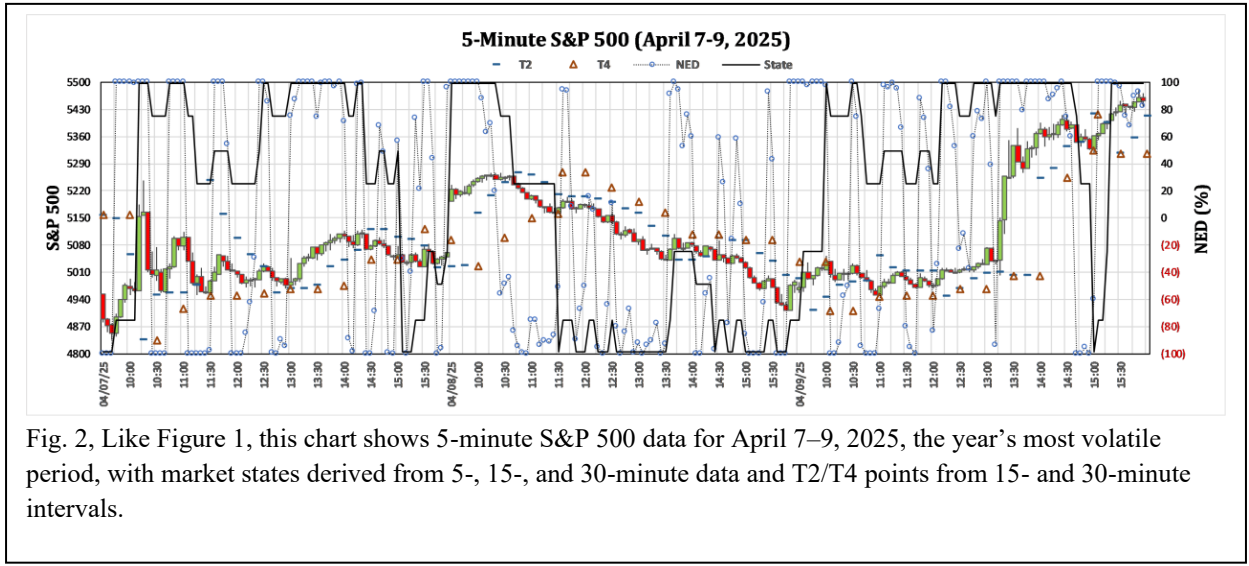

Fig. 2, Like Figure 1, this chart shows 5-minute S&P 500 data for April 7–9, 2025, the year's most volatile period, with market states derived from 5-, 15-, and 30-minute data and T2/T4 points from 15- and 30-minute intervals.

- **April 7**: The market opened at State 1 but spiked to State 8 after a social media rumor of a tariff pause. Although the White House denied it minutes later, optimism persisted, supported by late-day Signal 3s.
- **April 8**: A strong open quickly reversed as Signal 5s at the peak indicated insider selling. Successive Signal 4s drove the index from State 7 down to State 1 by midday, sustaining a decline into the close. The S&P ended near T2 and T4 upward turning points, setting up a rebound.
- **April 9**: At 9:37 a.m., Trump's post *"THIS IS A GREAT TIME TO BUY!!! DJT"* triggered a rapid surge from State 2 to State 8. Midday volatility reflected anxious swings between Signal 3s and 4s until Trump's 1 p.m. tariff suspension announcement drove a broad rally. A brief 3 p.m. dip to State 1 was countered by a strong Signal 3, enabling the market to close near its highs.

**Calm Intraday Price Patterns**

Forecasting price movements is especially difficult in low-volatility settings, where intraday volatility (V)—the ratio of the peak-to-trough range to the prior day's close—falls below 1%. In these periods, price behavior resembles white noise, lacking structure, autocorrelation, or momentum. This raises a key question: can any framework reliably identify minor local peaks and troughs? We show that, even under such constraints, meaningful price patterns can still be anticipated.

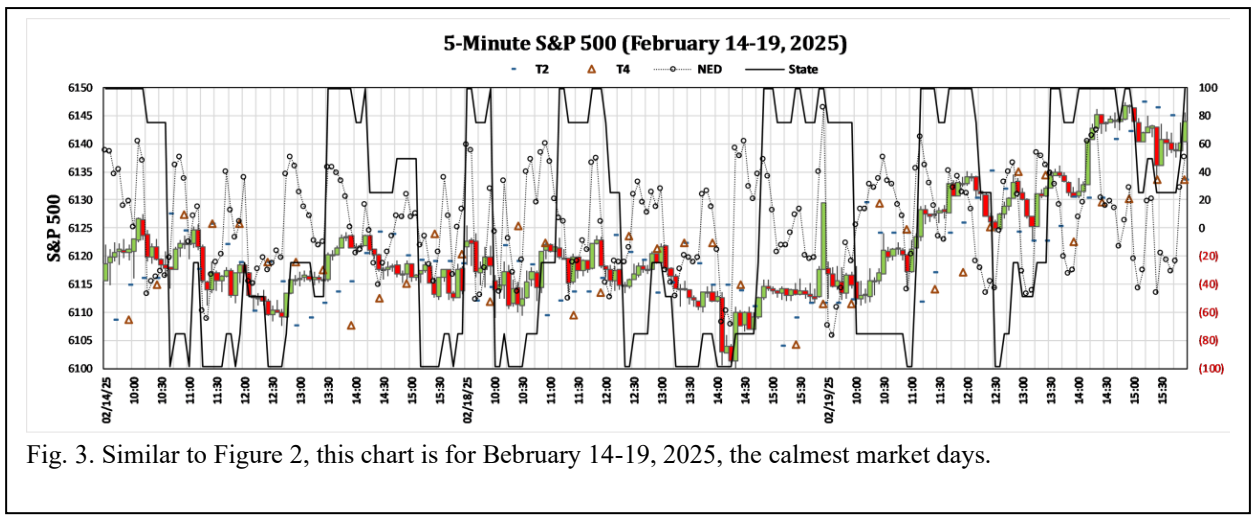

Fig. 3. Similar to Figure 2, this chart is for Bebruary 14-19, 2025, the calmest market days.

Figure 3 mirrors Figure 2 but covers the least volatile days of 2025. Peaks aligned with State 8 and troughs with State 1, while alternating Signals 3 and 4 drove sideways movements on February 14 and 18. A cluster of Signal 3s correctly forecast an upward trend on February 19. NED values stayed moderate, reflecting subdued sentiment typical of stable markets. However, predicting turning points proved less reliable under low volatility. For example, on February 14, a Signal 5 at 10:10 flagged selling at the peak, and convergence of T2/T4 with the S&P 500 signaled a decline at 10:40. Yet, an upward Signal 3 at 11:00 failed to anticipate the sudden drop from State 4 to State 1 at 11:10. Similar unpredicted declines occurred at 12:00 on February 14 and 13:00 on February 18. These misses highlight the limits of daily signals in calm markets, where subtle shifts require finer time resolution. To address this, we test a 1-minute chart to capture smaller, otherwise hidden, price changes.

Figure 4 presents the S&P 500's one-minute chart for the morning of February 14, highlighting the improved sensitivity of this finer resolution. It also shows NED values, market states (from 1-, 5-, and 15-minute data), and T2/T4 points (from 5- and 15-minute data).

Compared with the 5-minute chart in Figure 3, Figure 4 shows larger NED swings, reflecting greater sensitivity. Crucially, while Figure 3 gave no advance warning of the 11:10 and 12:00 downturns, Figure 4 displayed two Signal 4s at those exact points, correctly forecasting reversals. It also flagged stronger Signal 5s between 10:05 and 10:10 at the day's peak, indicating sustained selling

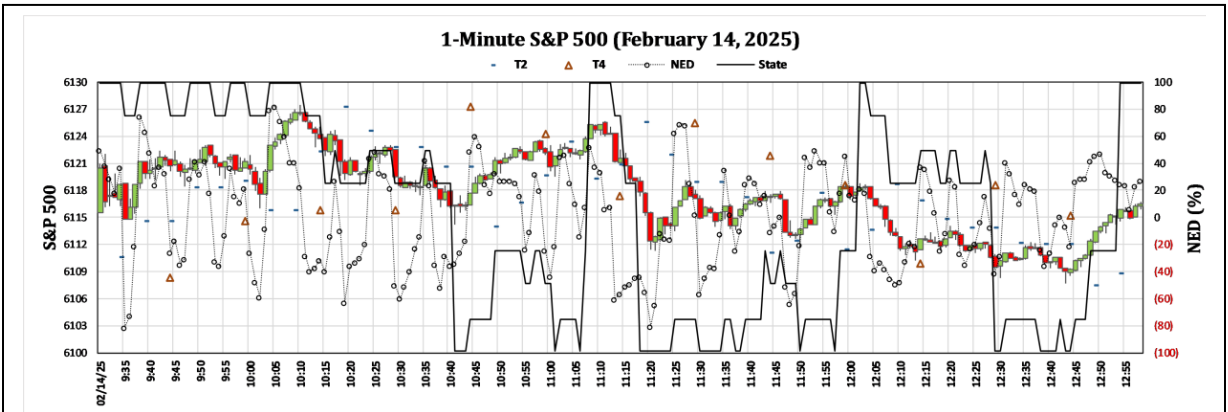

Fig. 4, This 1-minute chart of the S&P 500 (Feb 14, 2025, 9:30 AM - 1:00 PM) displays market states derived from 1, 5, and 15-minute data, with T2 and T4 turning points calculated from 5-minute and 15-minute intervals, respectively.

, and three Signal 4s between 10:10 and 10:40 that accurately predicted the decline—signals entirely absent in Figure 3. Notably, Figure 4 captures all the information in Figure 3 while adding finer predictive detail, showing temporal consistency across timeframes.

**Temporal Compatibility of Predictions**

The principle of temporal compatibility holds consistently across all timescales: forecasts derived from finer temporal resolutions necessarily embed the signals evident in broader, coarser timeframes, while simultaneously offering additional directional precision. An illustrative case is provided by the events of April 2025 (Figure 5). On the monthly chart, the S&P 500 transitioned from State 8 to State 5 during the height of tariff war turbulence on April 3rd. Although a strong Signal 3 suggested latent upward momentum, the monthly resolution alone was insufficient to determine whether a rebound would occur immediately or whether the index would continue its descent toward State 1. Greater granularity was therefore required. The daily chart (Figure 1) revealed that on April 3rd the market fell to State 1, triggered by Signal 4, with no immediate indication of reversal. A decisive Signal 3, implying a potential trend inflection, did not materialize until April 21st. To further discriminate between continued decline and the onset of recovery, it was necessary to consult the intraday chart (Figure 6), which captured the short-term fluctuations between April 3rd and April 7th.

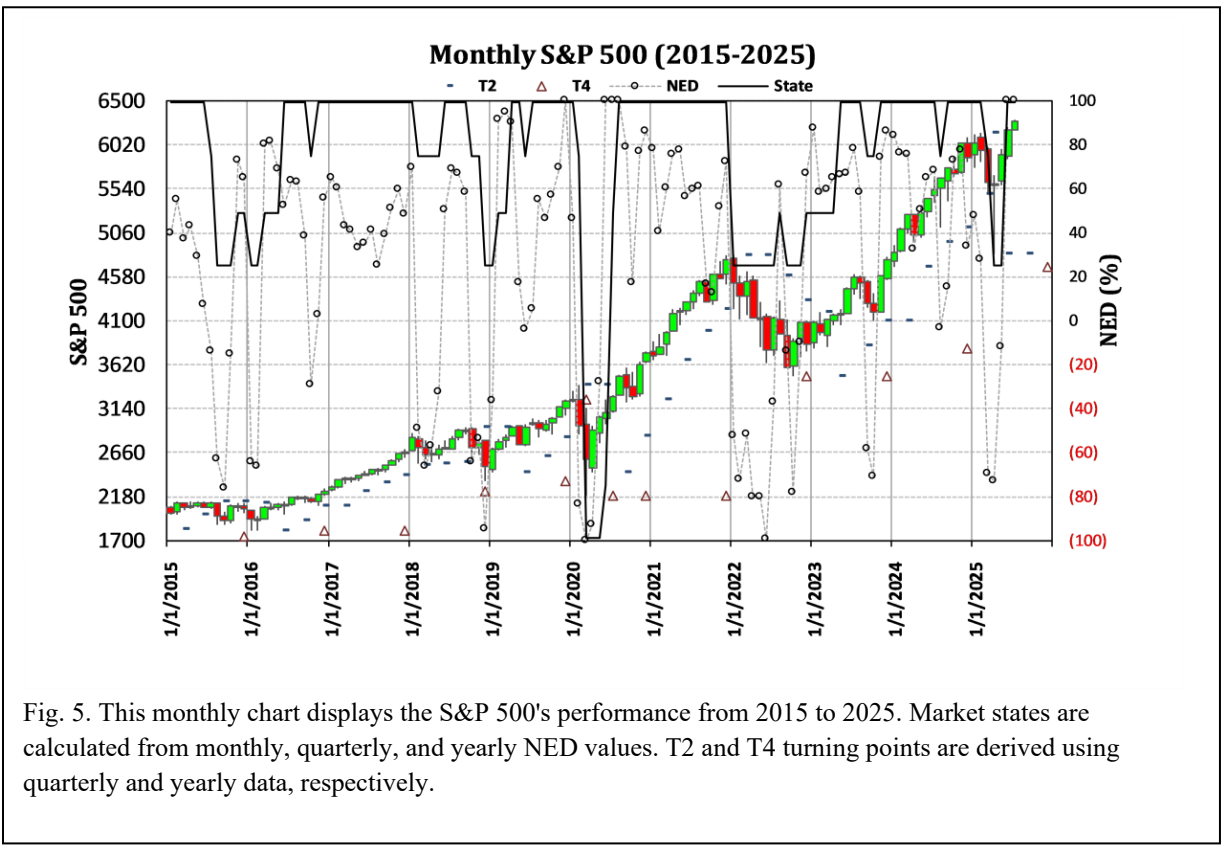

Fig. 5. This monthly chart displays the S&P 500's performance from 2015 to 2025. Market states are calculated from monthly, quarterly, and yearly NED values. T2 and T4 turning points are derived using quarterly and yearly data, respectively.

Figure 6 corroborated the prevailing downward pressure, displaying multiple Signal 4 occurrences on both April 3rd and 4th. A shift in market direction did not occur until April 7th, when rumors of a 90-day tariff delay emerged (as also reflected in Figure 2). This underscores that finer temporal resolutions generate more frequent signals, thereby enhancing the capacity to anticipate market movements.

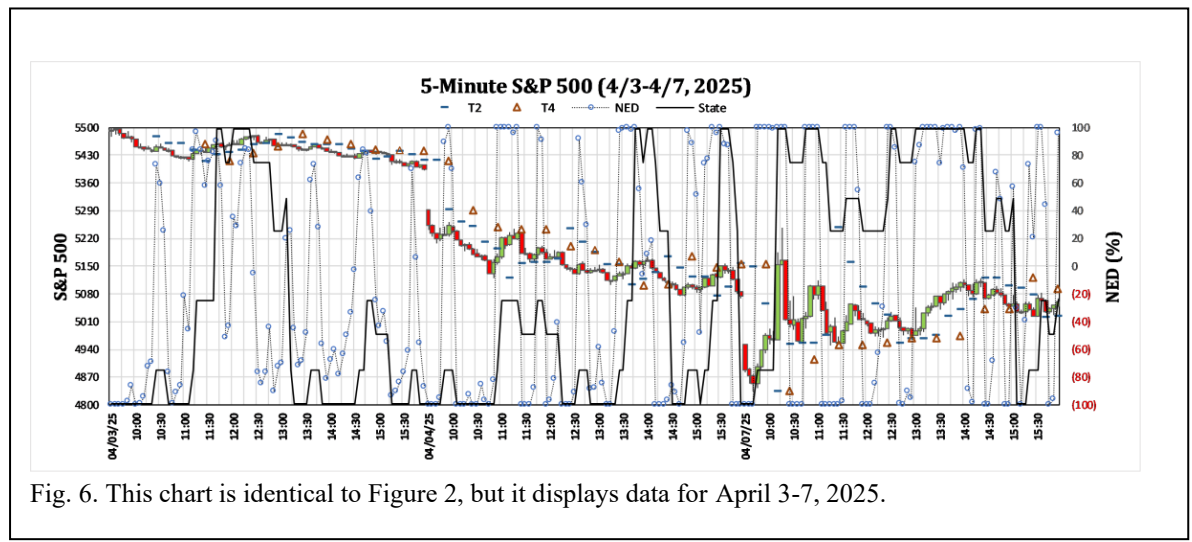

Fig. 6. This chart is identical to Figure 2, but it displays data for April 3-7, 2025.

By contrast, signals derived from coarser timeframes typically carry greater interpretive weight. For example, while the daily chart (Figure 1) indicated that the S&P 500 reached State 1 on April 3rd for the third time since March, the weekly chart (Figure 7) identified this as the first instance of State 1 since 2023—signifying a trough on a broader timescale and producing a strong Signal 3. Furthermore, although the daily chart captured persistent downward pressure, with multiple Signal 4 occurrences and repeated dips to State 1 throughout late March, it did not convey the larger-scale vulnerability that was evident in the weekly chart. As shown in Figure 7, the S&P 500 concluded March precisely at the downward T4 threshold (5489), poised to transition from State 6 to State 1. This transition was catalyzed by President Donald Trump's April 2nd announcement of "Liberation Day." This example illustrates how news frequently operates as a trigger, activating dynamics already embedded within the market structure. Conversely, news can also reshape longer-term market expectations; however, such adjustments tend to unfold gradually and are not instantaneously or fully incorporated into prices.

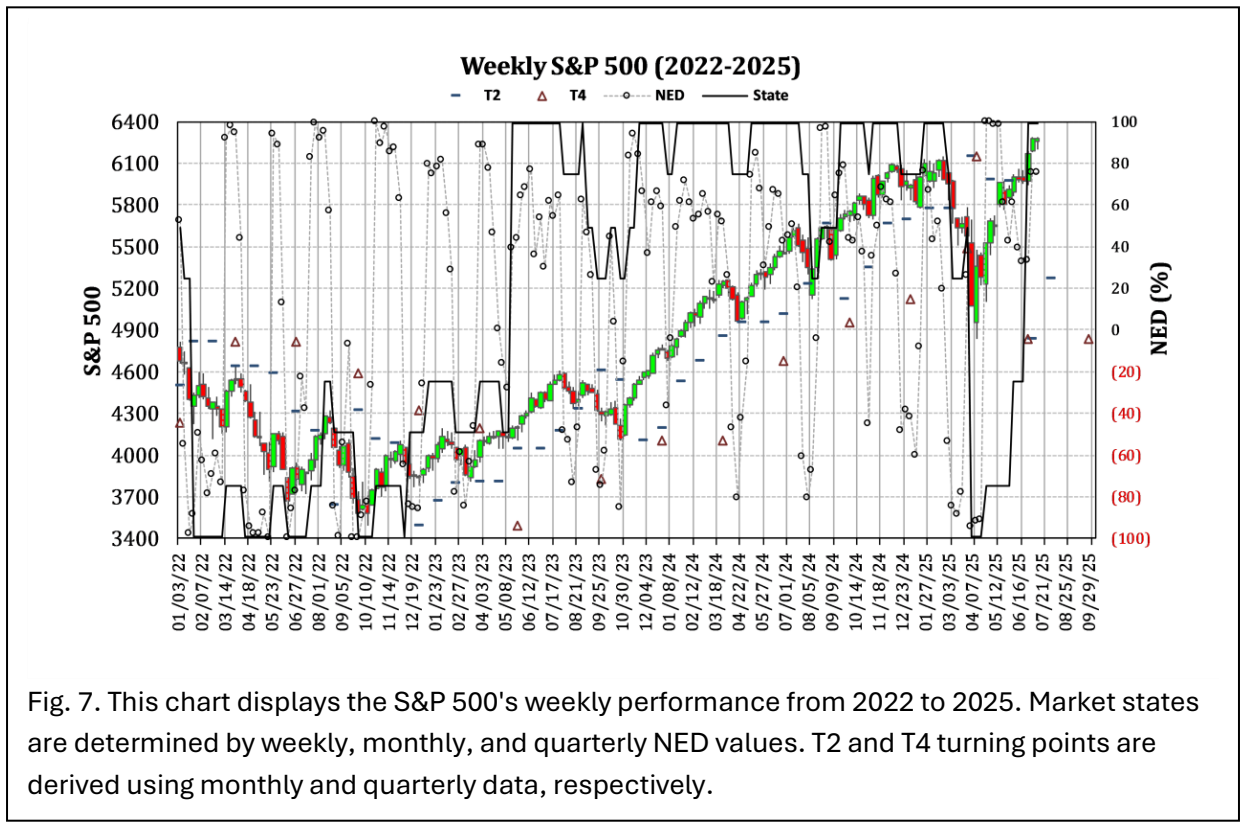

Fig. 7. This chart displays the S&P 500's weekly performance from 2022 to 2025. Market states are determined by weekly, monthly, and quarterly NED values. T2 and T4 turning points are derived using monthly and quarterly data, respectively.

**News Impact on Price Changes**

We have shown that endogenous market forces—such as the interplay between liquidity providers and demanders, as well as expectations across different time horizons—primarily govern price fluctuations at both short- and long-term scales. Nevertheless, external shocks, including political and economic news, exert substantial influence on investor expectations and behavior, as illustrated in the preceding section through several daily return cases from April 2025. In what follows, we demonstrate that such exogenous events can also impart long-lasting effects.

The price dynamics depicted in Figure 5 are generally predictable and interpretable through directional signals and T2/T4 state transition points. However, the onset of 2023 represented a notable deviation from these patterns. In December 2022, the S&P 500 reached the T4 downward transition threshold, accompanied by a strong Signal 4, both of which pointed clearly to an impending market downturn. This configuration was characteristic of a State 5–to–State 1 decline, reinforcing a bearish outlook that was further intensified by the seven interest rate hikes implemented during 2022.

Contrary to expectations, the market did not follow the projected downturn. It sustained State 5 for five months before advancing to State 8 in June, and the upward momentum continued following a strong Signal 3 in October. Although supportive economic developments—such as easing inflation and a resilient labor market—partly account for this transition from pessimism to optimism at the monthly scale, the principal catalyst was the rapid progress in artificial intelligence (AI).

On November 30, 2022, OpenAI officially released ChatGPT, which rapidly achieved viral adoption owing to its remarkable capacity to generate human-like responses, address complex questions, produce computer code, and assist across a wide range of tasks. Within weeks, it became one of the fastest-growing consumer applications in history, surpassing 100 million users by January 2023. This milestone ignited the AI boom of 2023, which served as a key driver of the equity market rally, particularly for mega-cap technology firms. Companies such as Nvidia, Microsoft, Google, Meta, Amazon, and Tesla experienced pronounced valuation gains in early 2023, much of which stemmed from progress in generative artificial intelligence and the growing belief in its commercial viability. The ensuing wave of 'AI-driven optimism' quickly evolved into a dominant market narrative, fueling the outperformance of technology-heavy indices such as the Nasdaq and the S&P 500. Investor enthusiasm centered especially on semiconductors, cloud computing, and enterprise technologies. As a result, U.S. equity markets performed unexpectedly well in the first half of 2023, defying widespread expectations of recession and rate-induced shocks. Over this period, the S&P 500 advanced 15.91%, while the Nasdaq surged by approximately 32%—its strongest first-half gain since 1983.

A major source of market unpredictability arises from the regular closures of equity markets. During the 17.5-hour daily pauses, as well as over weekends and holidays, breaking news can emerge, creating informational gaps that often translate into heightened volatility once trading resumes. In contrast, when news is released during active trading hours, its impact can be closely monitored, and market participants can adjust positions in real time. A clear illustration is provided by the events of April 9 (Figure 3). On that day, a major news release during trading hours drove the S&P 500 to rise by 8.7% in a single session—a movement that could be observed, processed, and acted upon as it unfolded. However, had the same announcement occurred during the overnight pause between April 8 and 9, such a dramatic increase would have been unpredictable, emerging instead as an abrupt discontinuity at the next day's opening. This discontinuous adjustment process highlights a structural asymmetry unique to financial markets. Whereas physical systems such as weather evolve continuously and can be measured without interruption, equity markets experience recurrent pauses that obscure the real-time integration of information.

Through a natural science approach, our findings suggest that market fluctuations, whether in the short or long term, arise from the interaction between exogenous shocks and endogenous market dynamics. Importantly, external shocks are not instantaneously and fully embedded in asset prices; rather, they are progressively incorporated into the collective expectations of investors, which in turn drive subsequent price movements. This interpretation provides a critical alternative to the Efficient Market Hypothesis (EMH), which posits that stock prices at any given moment reflect their 'intrinsic values,' adjusting instantly to the random arrival of new information. Yet this assumption raises a central question: how can market participants determine an 'intrinsic value' with precision in real time? As highlighted by Keynes (1936) and later by Fama (1989), the very notion of intrinsic value remains elusive and difficult to define consistently, even among experts.

**Discussion and Conclusion**

This paper demonstrates that the Extended Samuelson Model (ESM), grounded in a natural science approach, can uncover predictability in equity markets traditionally considered random. By integrating excess demand data (NED) across multiple timeframes, ESM successfully forecasts market states, directional price signals, and turning points, providing early warnings of major crises—including those of 1987, 2000, 2008, and 2020—and capturing intraday price movements previously deemed stochastic. The model's ability to track both liquidity takers and providers enables a dynamic, causal understanding of market evolution, moving beyond the correlational and binary limitations of conventional statistical early warning systems.

A fundamental source of market unpredictability lies in a feature absent from the physical sciences: the inherent discontinuity of trading. Unlike continuously evolving natural systems, financial markets operate in a punctuated cycle of closures. During these enforced pauses, significant news accumulates without being priced in. The market's reopening thus becomes a disruptive event of discrete price discovery, forcibly assimilating pent-up information and generating unpredictable volatility that has no direct analogue in physics or meteorology.

Even with frequent market closures, a natural science approach like the Extended Samuelson Model (ESM) can still anticipate short-term reversals, follow longer-term trends, and pick up subtle market shifts. ESM not only identifies peaks and troughs but also offers actionable signals on changes in market sentiment, effectively connecting fast intraday trading insights with broader crisis-level warnings. Its design ensures that finer timeframes capture all the information from broader ones, while also adding sharper directional guidance.

When markets trade continuously during intraday sessions, ESM shows reliable predictive power—even when major news breaks, as demonstrated in Figures 2, 3, 4, and 6. The framework's eight market states and six directional signals act like a trading compass, helping investors lock in profits as conditions shift. On even the calmest days, ESM can consistently capture 10-point moves in the S&P 500, regardless of direction. For E-mini S&P 500 futures, that translates to $500 profit per contract. This perspective helps explain the long-standing mystery of how Renaissance Technologies' Medallion Fund—under Jim Simons' leadership—was able to generate an extraordinary average annual return of 66% for more than 30 years without a single losing year. As Zuckerman (2019) describes, Simons categorized markets into eight states and capitalized on frequent intraday trading, a strategy closely aligned with the principles demonstrated by ESM.

The findings of the Extended Samuelson Model (ESM) demonstrate broad relevance beyond the S&P 500, applicable to individual stocks and international markets. This generalizability is possible because the ESM relies on a simple, realistic assumption: that price movements are fundamentally determined by the interaction between liquidity providers and demanders. To empirically test this generalizability, we applied the ESM to a broad cross-section of individual U.S. equities and to the foreign markets of Japan, China, and Hong Kong. The robustness of the findings across these disparate markets strongly suggests the ESM framework captures a universal dynamic of price formation. A comprehensive presentation of these validation studies and their consequences is reserved for a subsequent publication.

Looking forward, the application of ESM can extend beyond traditional equity markets. Its principles may be adapted for other asset classes, including commodities, foreign exchange, and digital assets, wherever real-time supply-demand imbalances can be measured. Moreover, the model opens avenues for integrating alternative data sources—such as social media sentiment, news analytics, and macroeconomic indicators—into predictive frameworks. By systematically quantifying market dynamics and agent interactions, ESM offers a promising path toward a science-based, proactive approach to financial stability and risk management.